\documentclass{article}
\usepackage[T1]{fontenc} 
\usepackage{textcomp}
\usepackage[utf8x]{inputenc} 

\usepackage{ismir,amsmath,cite,url}
\usepackage{graphicx}
\usepackage{color} 
\usepackage{multirow}
\usepackage{amsmath,amssymb}
\DeclareMathOperator*{\argmax}{arg\,max}

\usepackage{lineno} 

\title{Computational Analysis of Yaredawi YeZema Silt in Ethiopian Orthodox Tewahedo Church Chants} 
\multauthor
{Mequanent Argaw Muluneh$^{1,2}$ \hspace{1cm} Yan-Tsung Peng$^2$ \hspace{1cm} Li Su$^1$} { 
$^1$ Institute of Information Science, Academia Sinica, Taipei, Taiwan\\
$^2$ Department of Computer Science, National Chengchi University, Taipei, Taiwan\\
{\tt\small mequanenta@gmail.com, ytpeng@cs.nccu.edu.tw, lisu@iis.sinica.edu.tw}
}

\def\authorname{M. A. Muluneh, Y. T. Peng, and L. Su}

\begin{document}

\maketitle

\begin{abstract}
Despite its musicological, cultural, and religious significance, the Ethiopian Orthodox Tewahedo Church (EOTC) chant is relatively underrepresented in music research. Historical records, including manuscripts, research papers, and oral traditions, confirm Saint Yared's establishment of three canonical EOTC chanting modes during the 6th century. This paper attempts to investigate the EOTC chants using music information retrieval (MIR) techniques. Among the research questions regarding the analysis and understanding of EOTC chants, \emph{Yaredawi YeZema Silt}, namely the mode of chanting adhering to Saint Yared's standards, is of primary importance. Therefore, we consider the task of \emph{Yaredawi YeZema Silt} classification in EOTC chants by introducing a new dataset and showcasing a series of classification experiments for this task. Results show that using the distribution of stabilized pitch contours as the feature representation on a simple neural-network-based classifier becomes an effective solution. The musicological implications and insights of such results are further discussed through a comparative study with the previous ethnomusicology literature on EOTC chants. By making this dataset publicly accessible, our aim is to promote future exploration and analysis of EOTC chants and highlight potential directions for further research, thereby fostering a deeper understanding and preservation of this unique spiritual and cultural heritage. 
\end{abstract}
\section{Introduction}\label{sec:introduction}

The Ethiopian Orthodox Tewahedo Church (EOTC) chants 
hold immense cultural and religious significance in Ethiopia, yet they are largely overlooked 
\cite{Kebede:80}.\footnote{The Eritrean Orthodox Tewahedo Church, which separated from the EOTC administration system a few decades ago, also utilizes these chants. We acknowledge its important role in preserving this sacred form of church music.} 
The EOTC chant 
is believed to have originated with Saint Yared (505--571), who composed the three EOTC chanting 
modes (\emph{YeZema Siltoch} in Amharic language)
,\footnote{\emph{Siltoch} is the plural form of \emph{silt}. For simplicity, the Amharic phrase YeZema Silt and the English phrase chanting mode will be used interchangeably throughout this paper.} 
namely \emph{Ge}'\emph{ez},\footnote{The term \emph{Ge'ez} holds various connotations depending on context; here, it denotes one of the three chanting styles. Conversely, it also refers to the language and may have other applications.} \emph{Ezil} and \emph{Araray}. Saint Yared's pioneering musical compositions, liturgical chants, and associated dance movements had a significant impact on Ethiopian sacred music tradition \cite{Shelemay:92-debtera}. The \emph{Debterawoch} (also called \emph{Merigetawoch}), who are the expert musicians and heirs of Saint Yared, play a crucial role in the transmission and performance of the sacred music \cite{Kebede:80}. Ethiopian sacred music has been preserved through oral and written traditions, with written documents supporting and reinforcing the ongoing oral traditions \cite{shelemay1993oral}. 
The significance of the EOTC chants in Ethiopian culture and worldwide is evident through the two major spiritual mass celebrations that have been recognized by UNESCO as intangible cultural heritages: the Commemoration Feast of the Finding of the True Holy Cross of Christ (in 2013) and the Ethiopian Epiphany (in 2019).\footnote{\url{https://ich.unesco.org/en/state/ethiopia-ET?info=elements-on-the-lists}}
These two celebrations, primarily accompanied by the EOTC chants, are among the five intangible cultural heritages from Ethiopia registered by UNESCO. 

Despite its long history and development, the research of EOTC chants was quite rare. 
Among them, a renowned ethnomusicological work from Western academia was by Shelemay et al. \cite{shelemay1993oral,shelemay1993ethiopian}, based on the analysis of a series of EOTC chants collected in Addis Ababa, 1975. They discussed the oral and written tradition of EOTC chants, the EOTC chant music notation system, and further the definition of the three chanting modes, specifically the pitch sets used in each of the modes. 
It should be noted that in this work, all the recordings were transcribed and analyzed by ear. As stated in the paper, the analysis, for a limited number of recordings, was sometimes challenging when transcribing the non-Western music scales. With no indigenous classification of their pitch materials \cite{shelemay1993oral}, \emph{YeZema Siltoch} remains a primary research topic in the music theory of EOTC chants. 

This paper is a study on \emph{YeZema Siltoch} of the EOTC chants from computational perspectives. Our contributions in this paper are three-fold. First, we propose a new dataset for \emph{YeZema Silt} classification and analysis. 
Second, we benchmark the \emph{YeZema Silt} classification on the dataset using neural network classifiers with a number of features, 
primarily 
the pitch contour features which have been verified useful in analyzing various kinds of music \cite{RosenzweigSM19_StableF0_ISMIR,han2023finding,nikzat2022kdc,nadkarni2023exploring,caro2014creating,scherbaum2022tuning,Vidwans-SPCOM:2020}. Third, we perform a comparative study with \cite{shelemay1993oral,shelemay1993ethiopian} to echo, and to revise their statements as well: while the pitch sets used in \emph{Ezil} and \emph{Araray} was regarded as the same \cite{shelemay1993oral}, our numerical results indicate notable difference in between them. In the rest of this paper, we will have a background introduction of EOTC chants in Section \ref{sec:background}. The proposed dataset, benchmarks and the comparative study will be in Sections 3, 4, and 5, respectively. Conclusion and future works will be given in Section 6. 



\section{Background of EOTC chants}\label{sec:background}

\subsection{Features and Performance Traditions}\label{subsec:tradition}
The spiritual schools of the EOTC have several departments, locally known as \emph{Guba'e bet}. 
These departments include \emph{Nibab-bet} (reading practice), \emph{Zema-bet} (introductory to advanced level offices chanting), \emph{Qidase-bet} (or \emph{Kidase-bet}, liturgical chants), \emph{Qine-bet}\footnote{-\emph{ne}' is pronounced as in `Nelson'} (or \emph{Kine-bet}, poetry), \emph{Aquaquam-bet} (or \emph{Akuakuam-bet}, advanced chanting with accompaniments), and \emph{YeMetsahift Tirguame-bet} (exegesis of scriptures). 
The knowledge and skills acquired from each 
\emph{Guba'e bet} are crucial for 
understanding the chants. 
Each \emph{Guba'e bet}, which focused on chanting, has two or more slightly different vocal and performance styles \cite{mezmur-tsegaye-tourist:2011}. For example, \emph{Zema-bet} has \emph{Bethlehem}, \emph{Achabir}, \emph{Qoma}, and \emph{Tegulet}, and 
\emph{Qidase-bet} has \emph{Selelkula} and \emph{Debre Abay}. These nominations are based on the names of places where the center of excellence, that approves a senior student to be a teacher, is located. 
Such \emph{Guba'e bet}, for example, \emph{Bethlehem} has a slightly different vocal style, ornamentation, and notation complexity compared to \emph{Qoma}, and it also has its own swaying and religious dancing tradition with its own drumbeat.
The EOTC chants incorporate monophonic, antiphonal, and choral ritual performances. Our dataset is derived from  \emph{Qidase-bet}, which primarily focuses on monophonic and antiphonal ritual performance components without accompaniments. In contrast,  \emph{Aquaquam-bet} emphasizes religious dance and movements, primarily choral with some monophonic and antiphonal components. It is accompanied by prayer staffs known as  \emph{mequamia}, drums, and sistrums \cite{Baye2023:aquaquamZema,mezmur-tsegaye-tourist:2011}. The content of the chants - the text, whether poetic or unpoetic, is directly or indirectly based on the Holy Bible. 
The lyrics primarily employ Ge'ez (\includegraphics[scale=0.2,]{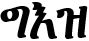}), an ancient Semitic language with a distinct script known as Fidäl. These chants play an essential role in the religious practices of nearly 43.5\% of the country's population, or over 32 million Orthodox Tewahedo Christians, according to the 2007 national census \cite{census2007}.\footnote{The Ethiopian and Eritrean faithful worldwide served by the chants is additional to the data reported in \cite{census2007}.}


The social groups involved in the chants include priests, deacons, and laypeople who attend the service hours. Traditionally, the chants were transmitted orally, with singers memorizing a repertoire of phrases and melodies to perform during liturgical celebrations. Several decades ago, chant manuscripts were handwritten on parchment, which refers to processed goat or sheep skins. Even today, some scholars adhere to this practice to uphold the church's cultural traditions. However, in recent decades, transmission has been supported by printed manuscripts for training along with oral traditions for actual performance.

Despite its rich heritage, the tradition of EOTC chants faces significant challenges. Many training centers are closing down due to absence of government support, insufficient community support for students \cite{mezmur-tsegaye-tourist:2011}, and the dominance of modern education since the 20th century. 
Despite the contributions of printing and recording advancements, 
the computational contribution to the Ge'ez language and the chants remains underdeveloped. 
Except for a few works on MIR \cite{Baye2023:aquaquamZema} and music generation \cite{GirmaZemedu:Hymn-Synthesis}, computational research on the EOTC chants is not as developed as it is for some other secular music. These issues highlight the need for more research on the EOTC chants. Our research aims to contribute to MIR-related tasks on the EOTC chants, addressing this gap.

\subsection{\emph{YeZema Siltoch} - Chanting Modes}\label{subsec:siltoch}

The EOTC chants encompass three primary \emph{YeZema Siltoch} (modes): \emph{Ge'ez}, \emph{Ezil}, and \emph{Araray}. They are typically employed sequentially or intermixed, sometimes aligning with the church calendar's seasons. Notably, during fasting periods, the \emph{Ge'ez} and \emph{Araray} modes predominate, while the \emph{Ezil} mode mostly reserved for holidays. These modes serve as conduits for conveying distinct emotions and seasonal themes within the EOTC chants \cite{Kebede:80}.

\begin{itemize}
\setlength\itemsep{0em}
\item \textbf{Ge'ez}: Characterized by a foundational, low tone, \emph{Ge'ez} chanting evokes a sense of despondency and solemnity. Rendered in a relaxed, subdued manner devoid of rhythmic constraints, it encapsulates feelings of despair, disappointment, and sorrow \cite{Kebede:80}. 
In \cite{shelemay1993oral}, the Ge’ez mode is interpreted as a \emph{chain of third} (\emph{a-c$^\prime$-e$^\prime$}) with ``chromatic auxiliary notes around the outer fifth'' ($\sharp$\emph{g}/ $\flat$\emph{b} around \emph{a}, and $\sharp$\emph{d$^\prime$}/ \emph{f$^\prime$} around \emph{e$^\prime$}).

\item \textbf{Ezil}: Positioned within a mid-range vocal register, the \emph{Ezil} (or \emph{Izil}) mode assumes a secondary role, characterized by its unassuming, moderate cadence. Emotionally neutral in essence, it is seldom utilized during fasting periods, maintaining a comfortable, ordinary vocal expression. Shelemay et al.  \cite{shelemay1993oral} stated that ``\emph{Ezil} uses the same pitch set as in \emph{Araray},'' but this pitch set is rendered as either \emph{c-d-e-g-a} or \emph{c-d-f-g-a}, implying that the third note lies in between \emph{e} and \emph{f} and causes ambiguity for Western ears. 

\item \textbf{Araray}: Distinguished by its high-pitched rendition, embellished with ornamental flourishes and a brisk tempo, the \emph{Araray} mode exudes vitality and jubilation. It serves as a vehicle for conveying animated expressions, elation, and manifestations of compassion, happiness and fulfillment. 
\end{itemize}

\begin{figure}[t]    
\centerline{\includegraphics [width=\columnwidth]{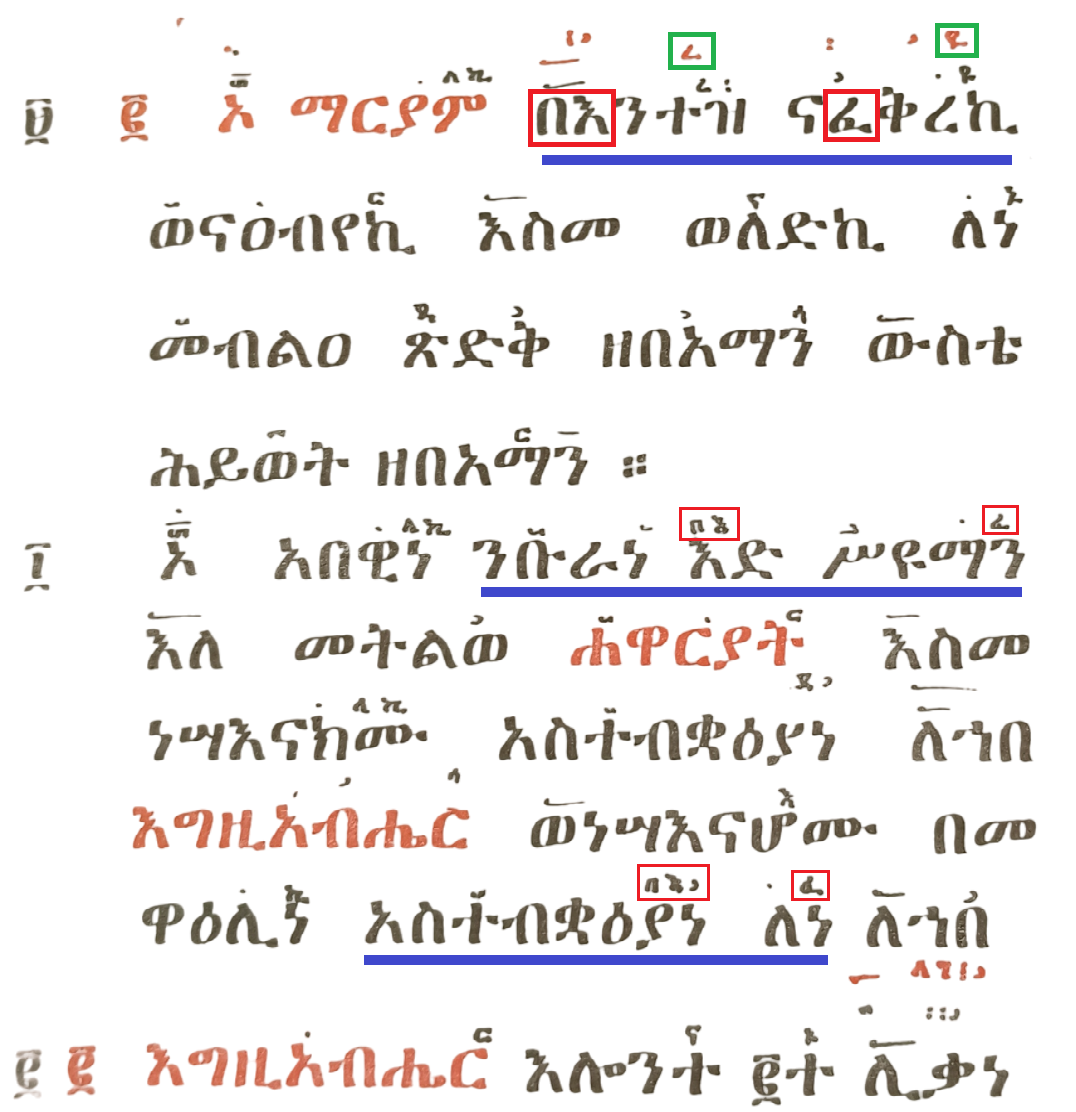}}
     \caption{Interlinear letter-based notations with interspersed neumes. 
  From the first underlined two words, the letters enclosed in red rectangles are used as short-form representations of the melody to be used over the other underlined words, sung with the same melody. 
     } 
     \label{fig:letter-notations} 
\end{figure}

The EOTC chants rely on a sophisticated system of interlinear notations, encompassing neumatic signs interspersed between letter-based representations \cite{shelemay1993oral, Kebede:80}. This notation system serves as the cornerstone of melodic expression in chanting. Although some notations are common across different chanting modes, they produce distinct melodies depending on the mode, making it challenging to identify a specific mode solely based on notation. Figure \ref{fig:letter-notations} provides an example of the notation system used in the EOTC chants. 

\section{Dataset}\label{sec:dataset} 

The dataset was manually collected from the \emph{Eat the Book} website,\footnote {\url{https://eathebook.org/}, We acknowledge the website's administrators for their invaluable contribution.} a hub of numerous audio books for most of the teachings in the EOTC school departments, with full and partial coverage. 
From the available audio books
, we selected the \emph{Se'atat Zema} (Horologium chant), which is part of \emph{Qidase-bet} department. All the audios selected for our dataset were recorded by a single scholar at a sampling rate of 44,100 Hz and in stereo channel. 



Our first step in the audio arrangement process involved narrowing the gap between the longest and shortest duration among the audios. 
Long audios, such as those over 13 minutes, were segmented into shorter audios of less than three minutes (180 seconds) 
in a way that preserves meaningful segments. 
This segmentation process also applied to audios that contained multiple chanting modes. For example, if an audio had 160 seconds of \emph{Araray} mode followed by 22 seconds of \emph{Ezil} mode content, it would be segmented into two separate audios of 160 seconds and 22 seconds. Recordings that were less than three minutes but still had multiple modes were also segmented based on the respective duration of the included chanting modes.

\begin{figure}[t]    
\centerline{\includegraphics[width=0.9\columnwidth, trim={0.2cm 0 0 0.25cm}, clip]{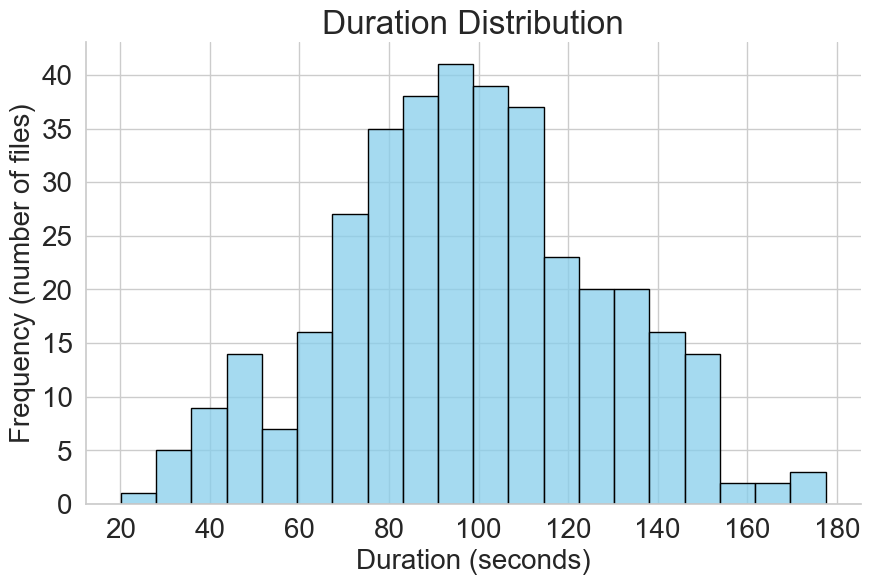}}
     \caption{Distribution of audio recording length (in secs).} 
     \label{fig:chant_durations}
\end{figure}

On the other hand, short audios, like a 19-second audio, were merged with neighboring context audios when applicable to our assumptions. 
If no neighboring audio with the same mode was found, it would be counted as a separate audio.
As we arranged all audios to be in a single mode, we have a corresponding mode label for each audio. 
Another audio cleaning process was removing non-chant segments as the recordings included short explanatory statements about the corresponding chants. We manually removed them to ensure that the full audio content will be for chanting. 
In this process we also have shortened the duration of significant silent regions, resulted in only two silent regions above two seconds, particularly 2.25 and 2.14 seconds. 
After such cleaning procedures, the overall duration distribution of our dataset, ranging from 20.142 seconds to 177.476 seconds, is shown in Figure \ref{fig:chant_durations}. 
As our immediate future work, we are working on expanding our dataset by including annotation of word-level lyrics to audio alignment as well as other features, which are not uncovered in this paper to keep the focus. We will do more research regarding other possible additional features. 

Table \ref{tab:DataDistribution} presents the comparison between the previously used dataset (i.e., the recordings collected by Shelemay \emph{et al.} in \cite{shelemay1993ethiopian}) and our dataset. 
While the previous dataset, with a total of 24 instances, is less than one hour, our dataset accounts for more than 10 hours, with a total of 369 instances. 
The chanting mode annotations of the dataset are available on \url{https://github.com/mequanent/ChantingModeClassification.git}.

\begin{table}
 \begin{center}
 \begin{tabular}{c|cc|cc} \hline
 & \multicolumn{2}{c|}{Shelemay and Jeffery \cite{shelemay1993ethiopian}} & \multicolumn{2}{c}{This work} \\\cline{2-5} 
\textbf{Modes} & \# & Length & \# & Length \\\hline
 Araray& 8 & 11m36s& 118 & 192m36s  \\
 Ezil & 6 & 9m56s& 176 & 291m29s \\
 Ge'ez & 10 &21m12s & 75 & 118m6s \\ \hline
 Total & 24 & 42m44s& 369 &  602m11s \\ \hline  
 \end{tabular}
\end{center}
 \caption{Data distribution among the chanting modes.}
 \label{tab:DataDistribution}
\end{table}

\section{Yaredawi YeZema Silt classification}\label{sec:classification}

As a preliminary study, we only consider using time-averaged audio features (i.e., the features ignoring the information lying in the temporal dimension) for \emph{Yaredawi YeZema Silt}  classification. Focusing on such features also supports our subsequent discussion on the pitch distributions of different chanting modes (see Section \ref{sec:analysis}). 

\subsection{Feature Representations and Classifiers}
Following previous works on the analysis of various kinds of music \cite{RosenzweigSM19_StableF0_ISMIR,han2023finding,nikzat2022kdc,nadkarni2023exploring,caro2014creating,scherbaum2022tuning,Vidwans-SPCOM:2020}, we consider pitch distribution, the distribution of the frame-level pitch values, for the classification task. 
Our pipeline of feature extraction mostly resembles \cite{scherbaum2022tuning,rosenzweig2022computer}, by having the stages of pitch contour extraction, stable region extraction, and pitch drift calibration. 
First, the pitch detection algorithm pYIN \cite{pyin} is utilized for pitch contour extraction. It sets the time resolution to 128 samples (5.8 ms) while the frequency resolution to 10 cents. After having the pitch contour, the pitch distribution is obtained by having a histogram over the frame-level pitch values with a frequency resolution of also 10 cents. To analyze the time-averaged aspects of the chanting modes, extracting the stable regions of the pitch contour while discarding the sliding, ornamental or other unstable components might be helpful. We therefore re-implement two stable region extraction methods, namely the \emph{morphetic} method and the \emph{masking} method, both proposed in \cite{RosenzweigSM19_StableF0_ISMIR}. 
There is also observable pitch drift during the performance. 
With an investigation of the data, we found that the pitch drifting along the whole recording is relatively small (around 1 semitone upward for the whole recording), 
so the pitch calibration process can be done straightforwardly with a linear regression. More specifically, the regression is performed on the pitch values 1 semitones around the global maximum of the pitch histogram. With the regression line with slope $s$, the pitch contour $f[t]$ indexed by time $t$ is calibrated to $f_{\text{calibrated}}[t]$ by having $f_\text{calibrated}[t] := f[t]-st$. 

\begin{table*}[t]
    \centering
    \begin{tabular}{c|c|c||llll|llll} \hline
           \multicolumn{3}{c||}{Feature representation} &   \multicolumn{4}{c|}{Within-dataset (5-fold CV)} & \multicolumn{4}{c}{Cross-dataset} \\  \hline\hline
\multirow{7}{*}{Pitch contour} & Calibration & Stabilization & full & 20 sec & 10 sec & 5 sec & full & 20 sec  & 10 sec & 5 sec \\ \cline{2-11}
&\multirow{3}{*}{No} & No & 96.20
& 91.51& 87.93 & 81.60 & 87.50 & 82.98 &  72.96  & 74.76 \\ 
& & Morphetic & 95.13
& 87.23&  83.47 & 73.35 &  83.33 & 84.40  & 76.67 & 70.97 \\ 
& & Masking & 94.85
&  83.85& 76.85& 64.32  &  87.50  & 74.47 & 68.52  & 55.41 \\ \cline{2-11}
& \multirow{3}{*}{Yes} & No & 98.11
&  89.92& 88.02&  80.78 &  75.00  & 80.85 &  77.04  & 69.64 \\ 
& & Morphetic & 95.66
&  87.71& 80.39& 70.58  &  79.17  & 73.05 & 70.00 & 69.07  \\ 
& & Masking & 92.94& 84.05&76.32 & 63.22  &  83.33  & 78.01 & 79.63 & 62.43  \\ \hline\hline
\multicolumn{3}{c||}{Time-average chromagram} &68.01& 66.63 &62.28 & 55.93&62.50 & 50.43 & 42.68 & 45.33  \\
\multicolumn{3}{c||}{Time-average mel-spectrogram} &64.20& 59.20 &55.16&54.61 & 37.50 & 48.72 &50.41 &47.91  \\
\multicolumn{3}{c||}{Time-average MFCC} &68.52& 66.62 & 66.16 & 65.42  &37.50 &35.90 & 36.18 &39.17  \\
\hline
 \end{tabular}
    \caption{Results (classification accuracy, in \%) of Yaredawi YeZema Silt classification.} 
    \label{tab:results}
\end{table*}



\begin{table}[]
    \centering
    \setlength{\tabcolsep}{4pt} 
    \begin{tabular}{|c|ccc|c|c|ccc|}
    \multicolumn{4}{c}{5-fold CV} & \multicolumn{1}{c}{}& \multicolumn{4}{c}{Cross-dataset} \\
    \cline{1-4}\cline{6-9}
           &  G & E & A &&& G & E & A \\ \cline{1-4}\cline{6-9}
         G & \textbf{92.0} & 2.67 & 5.33 & & G &\textbf{100.0} &0.0 &0.0 \\
         E & 1.14  &\textbf{97.73} & 1.14 & & E &0.0&\textbf{83.33}& 16.67\\
         A & 2.54 & 2.54 & \textbf{94.92} & & A &0.0&25.0&\textbf{75.0} \\\cline{1-4}\cline{6-9}
    \end{tabular}
    \caption{Confusion matrices over the \emph{Ge'ez} (G), \emph{Ezil} (E) and \emph{Araray} (A) classes. The reported classifier is trained on calibrated pitch contour with masking stabilization.} 
    \label{tab:confusion_table}
\end{table}







The pitch distribution features are therefore based on the six types of pitch contours: three stabilization modes (no stabilization, stabilization with morphetic method, and stabilization with masking method) times two calibration modes (with and without calibration).
Besides, several 
audio features 
are also compared: 
time-average mel-spectrogram, mel frequency cepstral coefficient (MFCC), and chromagram. 
The melspectrogram and MFCC are extracted using the 
\texttt{torchaudio} package \cite{hwang2023torchaudio}, 
while the 
chromagram is extracted with the \texttt{librosa} package \cite{mcfee-librosa:2015}. The time-average features of them are obtained simply
by taking average over the time axis. 


For the classifiers, we utilize the M5 (0.5M) model architecture 
proposed in \cite{dai2016deep}. The model is a fully convolutional network containing only 1-D convolution layers, max pooling layers and a global average pooling layer. Such a design has 
small number of training parameters and can capture the invariance in data \cite{thickstun2018invariances}. While this network was taken for raw waveform, we adapt it to operate in the frequency domain 
regarding it as an operator invariant to pitch shifting. 
To customize the model to our extracted features, we changed the receptive fields in the first convolutional layer from 80 to 3 when running on the non-raw-audio features in our experiments. For all the experiments, we adopt the categorical cross entropy 
loss function, Adam 
optimizer, learning rate of 0.001, 
batch size of 32, and 50 epochs, due to model convergence. 

\subsection{Experiment Settings}




To observe how the characteristics of \emph{YeZema Silt} vary across different recordings, we consider both the within-dataset and cross-dataset experiments. For the within-dataset experiment, we perform 5-fold cross validation (CV) on the proposed dataset and report the average classification accuracy. For the cross-dataset case, the model is trained on the proposed dataset and then tested on the recordings performed by a chanter from a different 
chanting department, specifically Zema-bet, in different time and location \cite{shelemay1993ethiopian}. The recordings we used from \cite{shelemay1993ethiopian}, described in Table \ref{tab:DataDistribution}, have a sampling rate of 44100 Hz with stereo channel with 0.33 and 4.05 seconds of shortest and longest audio recordings, respectively.  Lastly, to examine the reasonable identifiable audio duration among the chanting modes and how the duration affects the performance, we consider four input durations, namely 5 seconds, 10 seconds, 20 seconds and full length.

\subsection{Results} \label{subsec:results}


Table \ref{tab:results} lists the classification accuracy of all the experiment settings. First, the results of full length audio show that all the pitch distribution greatly outperform other audio features by a gap of over 25 percentage points. Also, the pitch distribution is more robust than the 
other audio features in the cross-dataset scenario, with a 
performance drop by 7 to 23 percentage points. However, comparing the six pitch distributions, it is not clear which calibration or stabilization mode is better. The best accuracy over all in the CV scenario is the calibrated but non-stabilized pitch distribution, but the trend does not apply to the cross-dataset case. Besides, we observe that 1) pitch contour stabilization does help on the accuracy for most of the cases, 2) using stabilization tends to reduce the performance gap between within-dataset and cross-dataset scenarios, and 3) the masking method can reduce this gap better than the morphetic method does, though the morphetic method typically has better classification accuracy. 
Lastly, there is a clear trend that a longer input audio leads to a better performance. This implies that \emph{YeZema Silt} is a long-term, song-level music concept, while it can also be signified to some extend upon a 10- to 20-sec duration, which is around the duration of a set of music notation.
 
Table \ref{tab:confusion_table} shows two example confusion matrices for both the within-dataset and cross-dataset cases. For the within-dataset case, the accuracy of each class basically follows the amount of data (\emph{Ezil} $>$ \emph{Araray} $>$ \emph{Ge'ez}, see Table \ref{tab:DataDistribution}). The trend is different for the cross-dataset case: all classification errors occur between \emph{Ezil} and \emph{Araray}, a result being in line with the experience of analysis \cite{shelemay1993oral}.

\begin{figure}[t]
    \centering
    \includegraphics[width=\columnwidth, trim={0.2cm 0cm 0.2cm 0cm},clip]{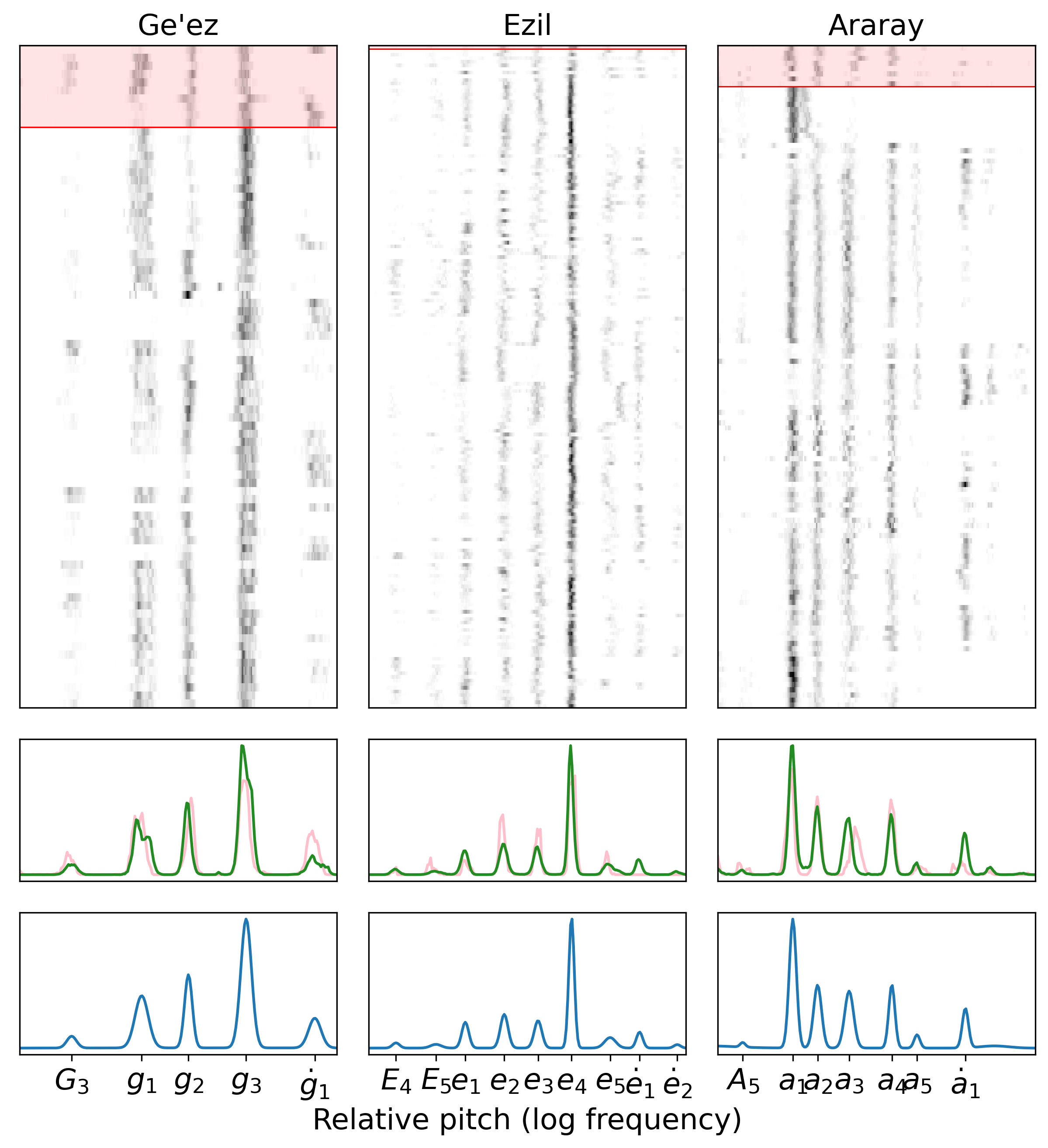}
    \caption{Illustration of pitch distributions for the three \emph{YeZema Siltoch}. Top: the aligned pitch distributions of all the recordings. A row in the 2-D illustration represents the pitch distribution of one recording. Darker color represents larger values. Red background represents pitch distributions of the recordings in \cite{shelemay1993ethiopian}. Middle: the average pitch distribution of the proposed dataset (green) and \cite{shelemay1993ethiopian} (red). Bottom: GMM-estimated pitch distributions for all the recordings from both datasets. The pitch value of each note name under the bottom row is listed in Table \ref{tab:gmm_result}.}
    \label{fig:pitch_distribution}
\end{figure}

\section{Analysis of Yaredawi YeZema Silt}
\label{sec:analysis}

The goal of our analysis of \emph{YeZema Silt} is using computational tools to individually identify the pitches utilized in the three chanting modes. 
Any attempt to this 
relies on some music theoretical assumptions. 
The classification results presented in Section \ref{subsec:results} supports two assumptions 
that facilitate the analysis
: first, \emph{YeZema Silt} is a song-level property that can be satisfactorily described with time-average pitch distributions; second, \emph{YeZema Silt} can be identified by a classifier invariant to pitch-shifting (i.e. convolution). 
On the other hand, the classification results also expose a few technical limitations. While the raw pitch distribution (i.e., without pitch contour stabilization) yields the best classification accuracy, it is highly noisy and therefore less applicable for our analysis purpose. In fact, we found in our study that the raw and the morphetic pitch distribution are relatively deficient in the below-mentioned analysis process. Therefore, instead of advocating a specific setting in terms of classification accuracy, we decided to use the calibrated pitch contour with masking stabilization method on the full length audio for subsequent analysis, although its performance is not the most favorable. It is worth noting that in this case, the performance gap between within-dataset CV and cross-dataset is relatively small among all settings. 

Our approach, which partly resembles \cite{scherbaum2022tuning}, contains three steps: 1) shift the pitch distributions of each recording such that each of them are best correlated (i.e., best aligned); 2) compute the average of the aligned pitch distribution for all the recording of the same chanting mode; 3) employ the Gaussian Mixture Model (GMM) to estimate the 
representative pitch set from the distribution. 

Specifically, the pitch distributions of two recordings $p_i$ and $p_j$ are aligned through pitch-shifting $p_j$ by $\xi_{ij}$ such that their cross-correlation $R_{ij}:=R_{ij}[\xi]$ is maximized:
\begin{equation}
    \xi_{ij}=- \xi_{ji}:=\argmax_\xi R_{ij}[\xi]\,.
\end{equation}
\indent The recording which has the highest average correlation with all the other recordings is considered as an anchor: the pitch distributions of all the other recordings are pitch-shifted to this  anchor according to their optimal $\xi$ and are then averaged for GMM fitting. 
The mean ($\mu$), variance ($\sigma^2$) and weight ($w$) of each GMM component then represents the pitch center, pitch variance and pitch weight. 
The GMM fitting process is initialized by user-specified mean values to enhance convergence \cite{scherbaum2022tuning}. To facilitate the discussion, only the components having variance smaller than 100 cents are considered as representative pitches.


The top row of Fig. \ref{fig:pitch_distribution} illustrates the aligned pitch distributions for the three chanting modes and the two datasets. We observe that the recordings in the same chanting modes typically have similar pitch distributions over the two datasets. Such a consistent trend is also observed from the average pitch distributions (middle row of Fig. 
\ref{fig:pitch_distribution}), which shows that only one pitch (the third peak from the left) from the two dataset in \emph{Araray} is somehow different. 

The bottom row of Fig. \ref{fig:pitch_distribution} shows the GMM-estimated pitch distributions for all the recordings from both datasets. By selecting the pitches summing up to maximal weights within one octave, we obtain three representative pitches for \emph{Ge'ez} (denoted as $g_1$, $g_2$ and $g_3$, from low to high), five for \emph{Ezil} (denoted as $e_1$, $e_2$, $e_3$, $e_4$ and $e_5$) and also five for \emph{Araray} (denoted as $a_1$, $a_2$, $a_3$, $a_4$ and $a_5$).\footnote{Here, the subscript number does not imply the hierarchical order of the musical scale (e.g., $g_1$ does not mean ``the tonic of the \emph{Ge'ez} mode''). The hierarchy of these pitches is another research question and will be considered as future work.} 
Other representative pitches outside this octave are also notated: the pitch being one octave below $g_3$ is denoted as $G_3$, while the pitch one octave above $g_1$ is denoted as $\dot{g_1}$. The same naming rules also apply for \emph{Ezil} and \emph{Araray}.

\begin{table}[t]
    \centering
    \begin{tabular}{c|ccccc}
    \hline
    Mode & Note name & $\mu$ & $\sigma^2$ & $w$ & $\Delta \mu$\\ \hline
    \multirow{5}{*}{Ge'ez} & $G_3$ & 361 & 11 & 0.034 & \multirow{2}{*}{486}\\
    & $g_1$ & 847 & 21 & 0.211 & \multirow{2}{*}{324} \\
    & $g_2$ & 1171 & 7 & 0.171 & \multirow{2}{*}{400} \\
    & $g_3$ & 1571 & 14 & 0.419 & \multirow{2}{*}{476} \\
    & $\dot{g_1}$ & 2047 & 18 & 0.112 \\ \hline
    \multirow{9}{*}{Ezil} & $E_4$ & 189 & 6 & 0.022 & \multirow{2}{*}{258} \\
    & $E_5$ & 447 & 14 & 0.023 & \multirow{2}{*}{223} \\
    & $e_1$ & 670 & 6 & 0.106 & \multirow{2}{*}{270}\\
    & $e_2$ & 940 & 7 & 0.151 & \multirow{2}{*}{234}\\
    & $e_3$ & 1174 & 7 & 0.123 & \multirow{2}{*}{232}\\
    & $e_4$ & 1406 & 3 & 0.416 & \multirow{2}{*}{268}\\
    & $e_5$ & 1674 & 15 & 0.068 & \multirow{2}{*}{204}\\
    & $\dot{e_1}$ & 1878 & 5 & 0.059 & \multirow{2}{*}{261}\\
    & $\dot{e_2}$ & 2139 & 5 & 0.013\\ \hline
    \multirow{8}{*}{Araray} & $A_5$ & 173 & 3 & 0.008 & \multirow{2}{*}{347} \\
    & $a_1$ & 520 & 6 & 0.318 & \multirow{2}{*}{172}\\
    & $a_2$ & 692 & 8 & 0.173 & \multirow{2}{*}{218} \\
    & $a_3$ & 910 & 10 & 0.176 & \multirow{2}{*}{297} \\
    & $a_4$ & 1207 & 5 & 0.134 & \multirow{2}{*}{174} \\
    & $a_5$ & 1381 & 4 & 0.027 & \multirow{2}{*}{335} \\
    & $\dot{a_1}$ & 1716 & 5 & 0.084 & \\ 
\hline
    \end{tabular}
    \caption{GMM-estimated mean ($\mu$, in cents), variance ($\sigma^2$, in cents), weight ($w$) of the representative note pitches in the Ge'ez, Ezil, and Araray modes. Reference pitch (0 cent) is 82.4 Hz. 
    The intervals (difference between two neighboring pitches, $\Delta \mu$) are listed in the last column.}
    \label{tab:gmm_result}
    \vspace{-0.3cm}
\end{table}

Table \ref{tab:gmm_result} shows the GMM-estimated parameters for the three modes. First, the pitches used in the Ge’ez mode are more flexible than other two modes, as can be observed by their variances than the pitches in other two modes. Among them, only $g_2$ has the variance less than 10 cents. $g_3$ and $g_2$ form a major third ($\Delta\mu=400$ cents) while $g_1$ and $g_2$ form approximately a minor third $g_2$ ($\Delta\mu=324$ cents). The pitches of $g_1$ and $g_3$ can vary by more or less semitones. 
Besides, we also observe that the octaves of $g_1$ and $g_3$ (i.e., $G_3$ and $\dot{g_1}$) also have large variances. This implies that such variance (flexibility of pitch) depend on the pitch name rather than the register.
These findings are basically in line with the statements (a scale $\sharp g$-$a$-$\flat b$-$c'$-$\sharp d'$-$e'$-$f'$ while $g$-$c'$-$e'$ are the stem pitches) made in \cite{shelemay1993ethiopian}. 

Both the \emph{Ezil} and \emph{Araray} modes have five representative pitches within one octave. However, the five representative pitches of them are different. For \emph{Ezil}, all the intervals lie between 200 cents (major second) and 300 cents (minor third), while for \emph{Araray}, the intervals distribute from 172 cents (less than a major second) to 347 cents (in between a minor third and a major third). In other words, there is a consistent trend that the intervals in \emph{Ezil} are more equally distributed than \emph{Araray}. There are also some flexible usage of pitch, for example, $e_5$ ($E_5$) in \emph{Ezil}. 
These suggest that the pitch sets found in \cite{shelemay1993ethiopian} needs revision: from our observation, each of the pitch sets used in the three EOTC chanting modes is distinctive. Besides, a mode is characterized by not only its pitch centers, but also its pitch variances. 

\section{Conclusion} \label{sec:Conclusion}
In this paper, we presented a research on a relatively underexplored music genre, the Ethiopian Orthodox Tewahedo Church (EOTC) chant, from three computational perspectives. First, through a rigorous data cleaning and annotation process, we presented a new and high-quality EOTC chant dataset, which can be extended for various music information retrieval (MIR) and music generation tasks. Second, we conducted a chanting mode (\emph{YeZema Silt}) recognition task using our dataset and achieved promising results. Additionally, this paper is, to our knowledge, the first to computationally analyze the pitch sets of the EOTC chanting modes, specifically \emph{YeZema Siltoch}, with new musicological insights. In the future, we plan to keep enriching the annotations of the datasets, by incorporating more details like lyrics, chanting options, reading tones and other potential features. Analyzing \emph{YeZema Siltoch} using the features in the temporal dimension and the new data annotations are also our ongoing projects.

The EOTC chants encompass a wide range of styles and forms. In this paper, we specifically concentrated on the \emph{Se'atat Zema} (Horologium chant), which falls under the \emph{Qidase-bet} department. Our objective is to encourage responsible research on EOTC chants, as computational research in this area can lead to technological advancements that enhance the learning process and increase accessibility. 
Diversifying the data and MIR of EOTC chants for the protection and promotion of this spiritual-cultural heritage is also our future work in the long term.



 



\section{Ethics Statement}
The chants we are working on belongs to the Ethiopian Orthodox Tewahedo Church (EOTC). The oral traditions and beliefs that mainly preserve these chants for more than 1500 years should be credited properly. 
This work aims to contribute on promoting the chants and finding solutions for their easy understanding. There is no intention to modify the chants in any form.

No explicit permission request is done to the Eathebook.org administrators as recordings are open to the public education and we are using for educational purpose. We value their great effort in making the chants publicly available.


\bibliography{ISMIRtemplate}

\end{document}